\documentclass[a4]{article}
\usepackage{graphicx}

\def\s{\,\,\,\,\,}
\def\mss{$\times 10^{-8}$ m/s$^2$\,}
\def\ug{$\mu$gal}

\title{\Large Precise Measurement of Gravity Variations During A
Total Solar Eclipse}

\author{Qian-shen Wang$^{1)}$,
Xin-she Yang$^{2)}$\thanks{{\large \bf Corresponding author}}, \\
Chuan-zhen Wu$^{1)}$
Hong-gang Guo$^{1)}$, Hong-chen Liu$^{1)}$, Chang-chai
Hua$^{1)}$ \\ \\
1) Institute of Geophysics, \, Chinese Academy of Sciences,
\, Beijing 100101,\, P R China. \\
2) Department of Applied Mathematics,
University of Leeds,  Leeds LS2 9JT, England.}

\begin{document}
\maketitle

\begin{abstract}

{\large The variations of gravity were measured with a high precision
LaCoste-Romberg D gravimeter during a total solar eclipse
to investigate the effect of solar eclipse on the
gravitational field. The observed
anomaly ($7.0 \pm 2.7$) \mss  during the eclipse
implies that there may be a shielding
property of gravitation. }

\end{abstract}


{\bf Citation Detail: }, {\it Phys. Rev. D} {\bf 62}, 041101(R), (2000). \\

\section{INTRODUCTION}

Although gravitation may has the property of shielding
in theories, it is very difficulty to test the
possible effect experimentally. If gravitation were carried
by particles, a mass between two bodies  could partially shield
each of them from the gravity of the other. Anomalies can be
expected in the motions of certain artificial Earth
satellites during eclipse seasons that behave like shielding of the Sun's
gravity as suggested by VanFlandern [1]. The possible existence of
gravitional shielding and gravitational-wave absorption [2]
and some theoretical analysis of a weak shielding of the gravitational
interaction by a disk of high temperature superconducting materials have been
investigated [3,4,5]. An experiment of electrically charged pendulum [6] was
carried out during an eclipse  to test the Saxl's effect [7] although
there was no noticeable effect observed.
Some related work were reviewed by Gillies [8].

If there were gravitational shielding, it would expect that
the effect shall be only significant during an eclipse
when gravity of the Sun may be shieldly slightly
by the moon so that the gravity on the Earth may fluctuate
accordingly, however such effect may be extremely small even if it would
exist. The present work was thus motivated
to test the possible effect of gravitational shielding
during the total solar eclipse with a high precision
modern gravimeter.

\section{EXPERIMENT}

To investigate the effect of possible gravitional shielding,
we conducted a precise measurement of the vertical
gravity variations during a total eclipse of the Sun on 9 March 1997 in China.
The observation and measurement during the total eclipse
were carried out in Moho,
Helongjiang province, China with the global position
$\phi=53^{o}29'20''$N and $\lambda=122^{o}20'30''$E,
which lies in the center of the shadow of the totality
during the eclipse. The parameters of the total eclipse
are: sunrise at 06:20:00 (local time), first contact
at 08:03:29, second contact at 09:08:18, third contact
at 09:11:04, and fourth contact at 10:19:50. The duration
of totality of the solar eclipse is 2 mins and 46 seconds.
The angular height of the Sun during the totality is
21$^{o}$.

A very high-accurate LaCoste-Romberg D gravimeter
(L \& R D-122) was used to measure the variations of
vertical gravitional acceleration with a high
precision of $2 \sim 3$ \mss or $2 \sim 3$ \ug.
The equipment was kept in a constant temperature
with $\pm 1^{o}$C inside an undisturbed room.
The output signal of the gravity variation from the
gravimeter was automatically collected by a PC.
The surrounding environment (within 200 meters)
was kept undisturbed during the whole process of
recording data so that there was no man-made gravitational
disturbance (e.g., gravity disturbance due to the
movement of people conducting the experiment).

The gravimeter (LaCoste-Romberg D) was very stable and had been used
for various field survey as well as daily record of tidal force for
several years. However, in order to ensure the accuracy of the measurement,
the gravimeter was installed well earlier before the eclipse. The gravimeter
reading was tested for several times to simulate real time recording.
The real-time recording began at 15:00 in the afternoon on 5 March 1997,
and go on continuously until 15:00 on 12 March 1997. The sampling reading
interval is 1 minutes. The sampling was increased near the eclipse.
The data reading was recorded at a rate of 2 reading every minute
from 06:00 am to 12:30pm and at a higher sampling rate of 1 reading
per second during the eclipse from 08:00am to 10:30am.

\section{RESULTS AND DISCUSSIONS}

The vertical gravitational acceleration measured consists
of several components: 1) gravitional forces due to
the Earth, the Sun and the Moon, and 2) the earth's
rotation. The former includes the static gravity
by the Earth and the tidal force by the Sun and the
Moon due to changes of moving positions. The tidal
component can be calculated theoretically with
a precision of 1 \ug \s or 1 \mss, which is a routine
practice in geophysics.

After making all these corrections,
the difference left shown Figure 1) is the variation of
vertical gravity during the eclipse due to some unknown
effect, which may be a possible shielding effect of
gravitation. The solid curve is the averaged values with a 10 minute
window and the variation can be more clearly identified.

The variation around zero has an amplitude
of $\pm 3 \sim 4$ \ug. The important and interesting
anomaly is that there exists two regions with significant
gravity decrease. One of such region occurred  within about
30 mins around 07:30am with a maximum significant decrease
of $6.0 \pm 2.5$ \ug, and another
took place within 30 mins around 10:20am with a maximum change of
$7.0 \pm 2.7$ \ug. The deviation is calculated by using the standard
formulae in measurement data processing. If the solid curve is used for
the calculation, the maximum changes shall be $5.3 \pm 1.4$ \ug
at fist contact and $6.8 \pm 1.4$ \ug at fourth contact,
respectively. These two changes took place
between first contact
and fourth contact, and quite closely related to
the timing of eclipse phases of first contact and
fourth (last) contact.

Figure 2 shows the measured gravity variation in the week of the
eclipse from 5 March 1997 to 12 March 1997. The significant variation
during the eclipse on 9 March 1997 is also shown (detail see Figure 1).
In plotting this figure, the data was averaged with a 10 minute moving window
so that the curve is more smooth than the actual measured data and the
signal looks  more significant. We can see that the reading was quite
stable before the eclipse and after the eclipse. The change during the eclipse is remarkable. Table I shows the number of data deviated from the
average value with a total of 10,080 data. Please note that the actual number of data during the eclipse is much more than those listed this table
(with a resampling rate of 1 reading per minute) because
the sampling rate during the eclipse  is much higher (1 reading per second).

The changes are quite significant and they are not
the effect of temperature and pressure changes.
According to the calibration precision of the LaCoste-Romberg
gravimeter provided by the manufacturer,
the variation of $8^{o}$C in temperature would lead
to 5 \ug change in gravity reading. The actual
temperature change in controlled room temperature during
the eclipse is within $\pm 1^{o}$C, so the actual
effect of temperature change is less than 1 \ug.
The actual change in pressure during eclipse
from 07:00am to 11:00am is about 1 mmH and the change
is less than 3 mmH in that whole day.
According to the manufacturer, the effect of
actual pressure change on gravity reading shall much
less than 1 \ug. Therefore, the actual noticeable changes
of gravity during the eclipse may imply
some extra-ordinary phenomenon associated with
gravity such as the possible shielding effect of
moon on the gravitational force of the Sun.
In addition, another puzzle is that the anomalies of the gravity
variations occurred at the first and last contact but not
during the totality. This certainly requires more precise
measurements in the future during totality of a solar eclipse.
\\

\def\tab{\hspace{2.5in}}
Table I: Measured Data Distribution \\
--------------------------------------------------------------- \\
Data deviation range (\ug)      \,\,\,\,\    Number of Data     \\
--------------------------------------------------------------- \\
$< $ 2                 \tab       9948 \\
$\ge$ 2    \,\,\,             \tab                 87 \\
$\ge$ 4    \,\,\,             \tab                 45 \\
--------------------------------------------------------------- \\

In summary, we have used the best available gravimeter,
with a high precision of 2 $\sim$ 3 \ug, to measure
the variation of vertical gravity during the total
eclipse on 9 March 1997. Although there was no noticable
changes around the totality during the solar  eclipse,
we have observed quite significant decrease in
vertical gravity during the first contact and the last
contact. The may imply the new property of gravitation,
which certainly needs more high precision experiments
to be conducted in the future especially during solar
eclipse. Although the purpose of this short
paper and the present work is not intended to prove
the shielding effect of gravitation, however, we would be
delighted if the present work can initiate more work
on the possible new property of gravitation.

{\bf Acknowledgement}: We would thank the referee(s) for their insightful
comments which has greatly improved the manuscript, especially for
the kind suggestion of averaging the data over the 10 minute interval.
The work was supported by the
National Natural Science Foundation of China. The authors
are grateful to the help from  the Moho geophysical station
of Chinese Academy of Sciences.

\section*{REFERENCES}

\begin{enumerate}
\item T. VanFlandern, {\em Astrophys. Space Sci.}, {\bf 244}, 249 (1996).

\item V. Desabbata and C. Sivaram, {\em Nuovo Cimento}, {\bf B 106},
      873 (1991).

\item  G. Modanese, {\em Europhys. Lett.}, {\bf 35}, 413 (1996).

\item  C. S. Unnikrishnan, {\em Physica}, {\bf C 266}, 133 (1996).

\item  E. Podkletnov and R. Nieminen, {\em Physica}, {\bf C 203}, 441 (1992).

\item Y. C. Liu, X. S. Yang, T. R. Guan et al., {\em Phys. Lett.},
{\bf A 244}, 1 (1998).

\item E. J. Saxl, {\em Nature}, {\bf 203}, 136 (1964).

\item G. T. Gillies, {\em Reports Prog. Phys.}, {\bf 60}, 151 (1997).
\end{enumerate}

\begin{figure}
\centerline{\includegraphics[width=5in]{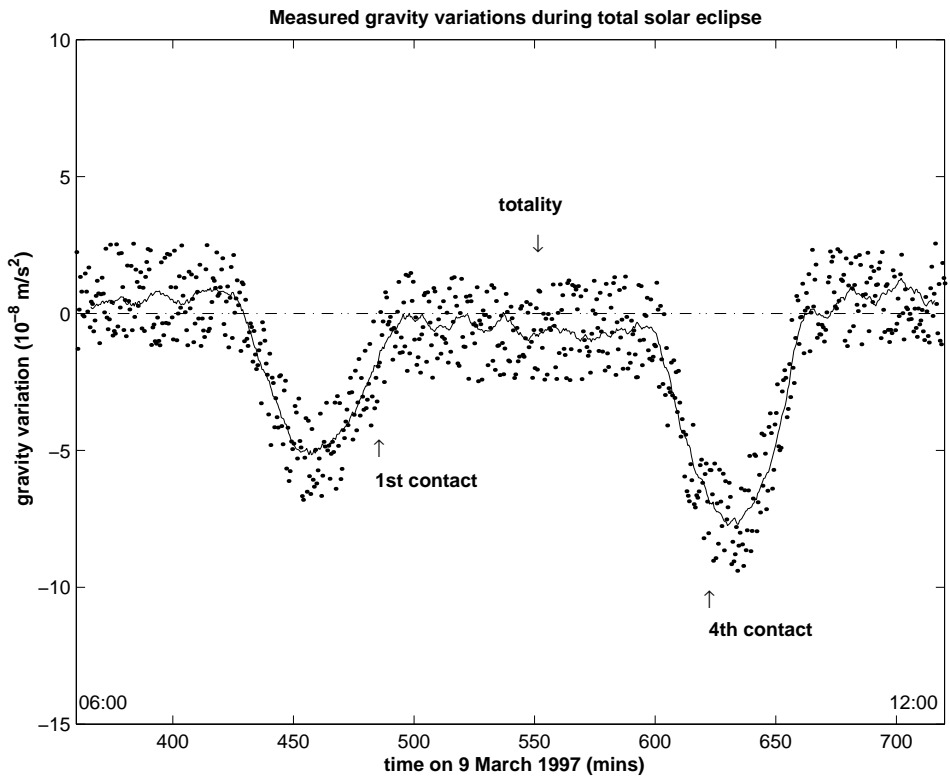}}

\caption{ Variations of vertical gravity measured during the
total solar eclipse on 9 March 1997. The solid curve is the averaged
variation over a moving 10-minute window.
Two regions of gravity anomaly during the
eclipse were observed, which may be the effect of
gravitational shielding.  }
\end{figure}

\begin{figure}
\centerline{\includegraphics[width=5in]{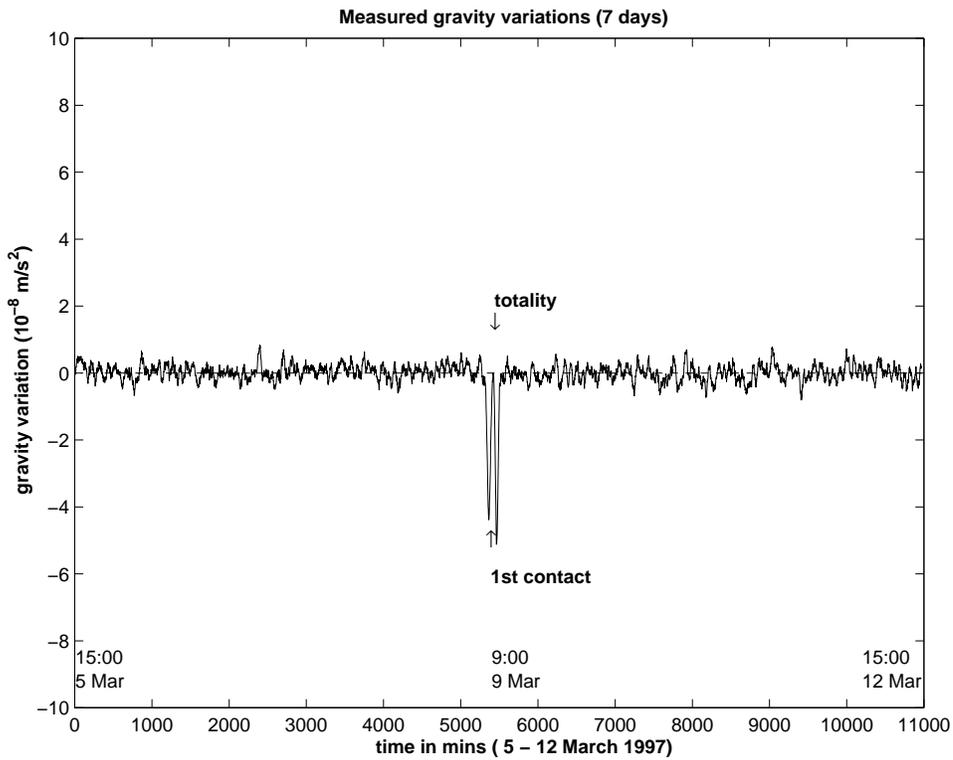}}

\caption{ Measured variations of vertical gravity measured during the whole week from 5 March to 12 March 1997. Significatn change was observed
during the eclipse on 9 March 1997, which is shown in more detail
in Figure 1. }
\end{figure}

\end{document}